# Dynamics and electrodynamics of moving real systems in the Theory of Reference Frames

Daniele Sasso

dgsasso@alice.it

**Abstract**

The purpose of the dynamics of moving systems is to search for the mathematical model that describes the link between the resultant applied force, that is the cause, and the speed of system that is the effect. This mathematical link provides the law of motion and from Newton forward it has been specified by an differential equation. Starting from the principles of the Theory of Reference Frames it is possible to arrive at new interesting results in the study and in the analysis of dynamical systems provided with linear symmetry and central symmetry. Really the most interesting consequences are relative to the behaviour of microphysical systems and elementary particles. Some kinematic and dynamic physical parameters, valid for analysis of mechanical systems, lose meaning in microphysical systems. For these systems it is necessary to define therefore new parameters on the ground of the behaviour of observable physical magnitudes. Surely the most important physical quantity, defined in this paper and relative to elementary particles, is the electrodynamic mass that is different from mass of mechanical systems. This parameter allows to understand some features of the behaviour of elementary particles in inertial and not inertial reference frames. In the last part of the paper an important relation between time and mass is proved and new general equations of space time are defined for any reference frame.

**Contents**

**Introduction**

The classical dynamics was elaborated in the sixteenth and the seventeenth century by some great scientists among whom Galileo Galilei and Isac Newton have a preeminent place. Galileo and Newton specified the principles of inertia, action, action and reaction, and relativity. At the beginning of the twentieth century by his new theory of relativity Einstein made a deep revision of classical dynamics. He introduced new kinematic definitions of space and time and the most interesting dynamic concept of Special Relativity[1][2] is certainly the variation of mass by velocity. In the second part of his fundamental scientific work (1905) Einstein discussed the theme regarding the dynamics of electron and he demonstrated the transversal mass and longitudinal mass of electron increase by velocity. He extended then these results to all elementary particles and also to whatever mass without charge. It is interesting now to see as the dynamic issue is discussed and solved within the limits of the Theory of Reference Frames. In order to attain this purpose it is opportune to distinguish the study of physical systems between systems with linear symmetry and systems with central symmetry. Systems with linear symmetry are systems that move with rectilineal motion determined by an applied force. Systems with central symmetry have a motion determined by a force field produced from a central pole and this motion is not necessarily rectilineal. We will distinguish then the study between mechanical systems and microphysical systems (elementary particles) and for microphysical systems we will introduce the concept of electrodynamic mass that has a different behaviour with respect to the mass of mechanical systems.
Before initiating this study it is appropriate to specify some basic definitions.
A physical system is a set of components linked among them in structural or functional way. Particularly a single component is a simple system and similarly a body with mass $m_o$ is a simple system. If a system under the action of a force gains a velocity only by reason of its mass then the considered system is a mechanical system. If a system under the action of a force gains a velocity only by reason of its charge the considered system is an electrodynamic system. If various forces act on the system the vector resulting force is applied to the centre of mass of the system.

## 1. Dynamics of systems with linear symmetry

### 1.1 The general law of motion for mechanical systems

In classical physics the general law of motion for a mechanical simple system with mass $m_o$ under the action of a force F is given by Newton's law

$$F = m_o \frac{dv(t)}{dt} \qquad (1.1)$$

If the force is constant and continuous (1.1) states the effect of a constant force on mass $m_o$ is a constant acceleration and therefore the consequent motion is an uniformly accelerated motion. In this way Newton did a deep revision of the Aristotelian physics which affirmed on the contrary the effect of a constant force on a mass was a constant speed. The Einsteinian relativity substantially confirmed Newton's equation and introduced the concept of variable mass with the velocity which Einstein demonstrated by Lorentz's transformations. Within the limits of the Theory of Reference Frames (TR) it is possible come to a new equation of motion for a system in real conditions [3][4].

A mechanical system with mass $m_o$, free and not tied, under the action of a constant and continuous force F acquires a motion and the equation of motion is determined in each instant by the dynamic equilibrium between the applied force F and the resistant force $F_r$.

Resistant force can be various but certainly there are in real conditions the inertial force $F_i$ (internal resistant force) and the resistant force $F_m$ of medium where the motion happens (external resistant force).

It is therefore

$$F_r = F_i + F_m \quad \text{and}$$
$$F = F_i + F_m$$

Inertial force is proportional by means of mass $m_o$ at the instantaneous acceleration of system according to Newton's equation and in opportune conditions resistant force of the medium is proportional at the instantaneous velocity of system by means of a constant coefficient K which is called "resistant coefficient of medium".

The general law of motion of a real mechanical system is therefore

$$F = m_o \frac{dv(t)}{dt} + K\, v(t) \qquad (1.2)$$

The (1.2) is a linear differential equation of first order and it is the mathematical model of the considered system. If the system is initially at rest with v(0)=0 for t=0, the solution of the differential equation is

$$v(t) = \frac{F}{K}\left(1 - e^{-kt/m_o}\right) \qquad (1.3)$$

where $V_o = F/K$ represents the terminal speed of system. Moreover the quantity $T = m_o/K$ is the " mechanical time constant ".

If we draw by a diagram the law of motion we have the graph of figure 1

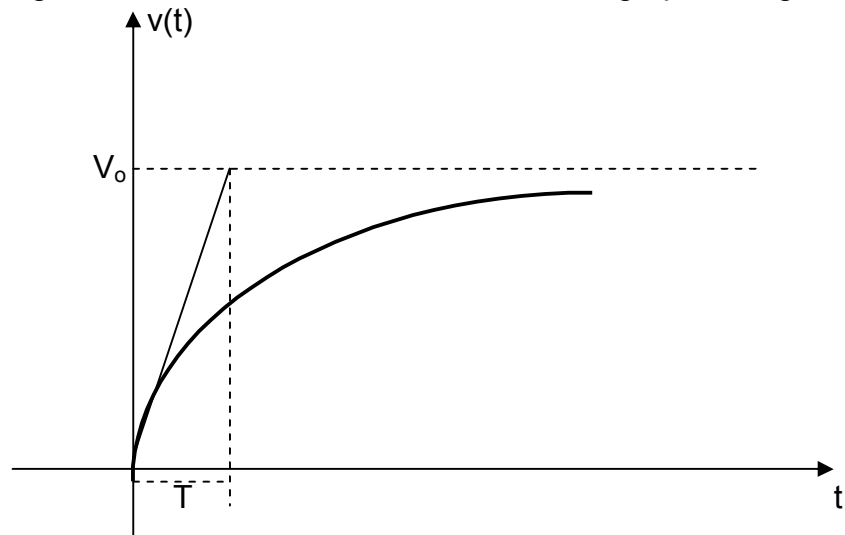

Figure 1. Velocity trend of a mass in real conditions under the action of a constant and continous force F.

Observing the graph we can see that in general the physical effect of a constant and continuous force, applied to a mechanical system with mass $m_o$, is in real conditions a variable acceleration and therefore the motion is a varied motion.
It is possible then, in first approximation, to replace (fig.2) the exponential real trend of the motion, indicated by the curve 1, with the curve 2 which has an increasing linear trend until the time T (initial instants of motion) and a constant trend after the time T.

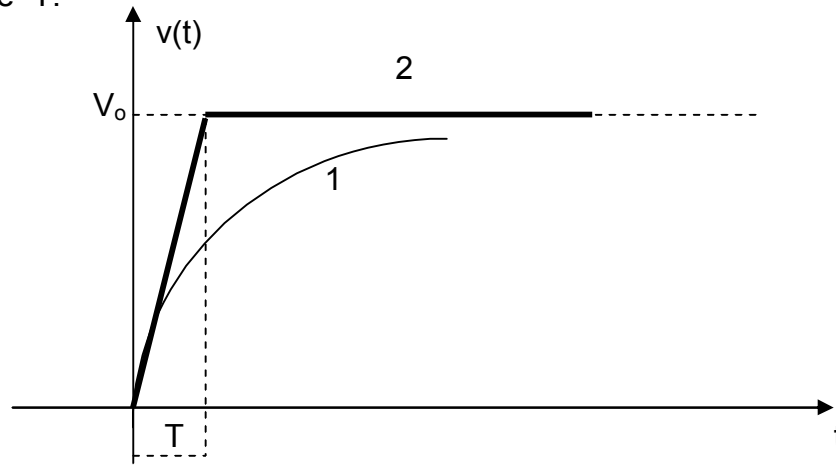

Figure 2. Approssimate trend of velocity (curve 2).

Examining the trend 2 in figure 2 it is possible to deduce, in first approximation, the effect of a constant and continuous force applied to a system with mass $m_o$ is a constant acceleration in initial instants of motion (Newtonian concept) and a constant speed in the following instants (Aristotelian concept).
The general law of motion, here demonstred, estableshes therefore a settlement of dispute between Aristotelian and Newtonian physical visions. It is possible to observe the speed of system depends on the mass only in acceleration whereas the terminal speed $V_o$ is always independent of the mass and dependent on the resistant coefficient k of medium.
Observing the exponential trend of the curve 1 and making simple calculations we can say the system spends a time about T'=5T to achieve pratically the terminal speed $V_o$ because after this time certainly the system exceeds the 99% of $V_o$.
If we examine now the physical state where the resistant coefficient k is zero or close to zero (k≈0), the general law of motion in these conditions is coinciding with Newton's law $F = m_o dv(t)/dt$. Solving the precedent equation by the condition with null initial speed, we have $v(t) = Ft/m_o$. Under the action of a constant and continuous force, in the absence of external resistant forces, the speed of system increases with linear law to acquire in theory an infinite speed. If at time $t_o$ the system has a speed $V_o$ and we suppose the applied force F becomes zero at this time, the law of motion for k=0 is

$$m_o \frac{dv(t)}{dt} = 0 \qquad (1.4)$$

The solution of this equation is $v(t) = \text{constant} = V_o$.
In these conditions (k=0) a mechanical system, with a constant initial speed and not placed under any action of external forces (F=0), is inclined to keep its state of constant rectilinear motion: this is the "inertial motion" that is at the root of the "Principle of inertia" of the classical physics. In our study the classical principle of inertia is valid only in ideal physical conditions (k=0) and not in real physical conditions. In fact in real conditions (k≠0) the law of motion with F=0 is

$$m_o \frac{dv(t)}{dt} + K\, v(t) = 0 \qquad (1.5)$$

and so the mechanical system doesn't keeps its state of motion ($v(t)=V_o e^{-t/T}$).

### 1.2 Energy considerations about motion

In order to move a mass $m_o$ initially at rest along an elementary distance ds, a constant and continuous force F executes an elementary work $dW=Fds$.
For the general law of motion we have

$$dW = m_o \frac{dv}{dt} ds + k\, v\, ds$$

$$dW = m_o\, v\, dv + k\, v^2\, dt$$

As the mechanical system spends in theory an infinite time (fig.1) to achieve the terminal speed $V_o$ the total work executed by the force F to lead the mechanical system to terminal speed is

$$W = m_o \int_0^{V_o} v\, dv + k \int_0^{\infty} v^2\, dt \qquad (1.6)$$

and prosecuting the calculation we have

$$W = \frac{1}{2} m_o V_o^2 + k \int_0^{\infty} V_o^2 \left(1 - e^{-t/T}\right)^2 dt$$

$$W = \frac{1}{2} m_o V_o^2 + \lim_{t \to \infty} E_k \qquad (1.7)$$

The term $E_c = m_o V_o^2/2$ is the kinetic energy acquired by the mechanical system.
The term $E_k$ is the energy that the force spends in the time t in order to win the external resistant force.
The work executed by a force F on a mechanical system with mass $m_o$ to lead it from the null initial speed to the terminal speed $V_o$ is subdivided between the variation of kinetic energy and the dissipated energy.
As the system spends about a time $T'=5T$ to achieve practically the terminal speed, considering the energetic balance only in this time we have

$$W = m_o \int_0^{V_o} v\, dv + k \int_0^{T'} v^2\, dt$$

$$W = \frac{1}{2} m_o V_o^2 + \frac{7}{2} m_o V_o^2 \qquad (1.8)$$

and so the work of the force F on the mechanical system in the time $(0,T')$ is subdivided between the variation of kinetic energy $E_c$ and the dissipated energy that is $E_k \approx 7E_c$. After the time T' the work carried out by F is used practically only to overcome external resistant force.

If the resistant coefficient k is null or practically null (k=0) we have $W = m_o V_o^2/2 = E_c$ and so all work executed by force F to lead the mechanical system to speed $V_o$ is converted in kinetic energy.

### 1.3 Transformation of the general law of motion for inertial systems.

The general law of motion in the reference frame at rest S[x,y,z,t,] is described by the equation (1.2)

$$F = m_o \frac{dv(t)}{dt} + K v(t)$$

With respect to the moving inertial reference frame S'[x',y',z',t'] for the principle of relativity the equation of motion is

$$F' = m_o' \frac{dv'(t')}{dt'} + K' v'(t') \qquad (1.9)$$

Supposing the reference frames S and S' are coincident at the initial time t=t'=0, supposing the moving frame S' moves with constant speed $v_u$ in the direction of increasing x and supposing the speed **v** is in the direction of increasing x (fig.3), the transformation equations from S to S' are[5][6]

| | |
|---|---|
| t' = t | (constancy of the inertial time) |
| x' = x – $v_u$ t | (transformation of the coordinate x) |
| y' = y | (constancy by supposition of the coordinate y) |
| z' = z | (constancy by supposition of the coordinate z) |
| $m_o'$ = $m_o$ | (constancy of mass in inertial systems) |
| F' = F | (constancy of force in inertial systems) |
| v' = v - $v_u$ | (vectorial composition of speeds) |

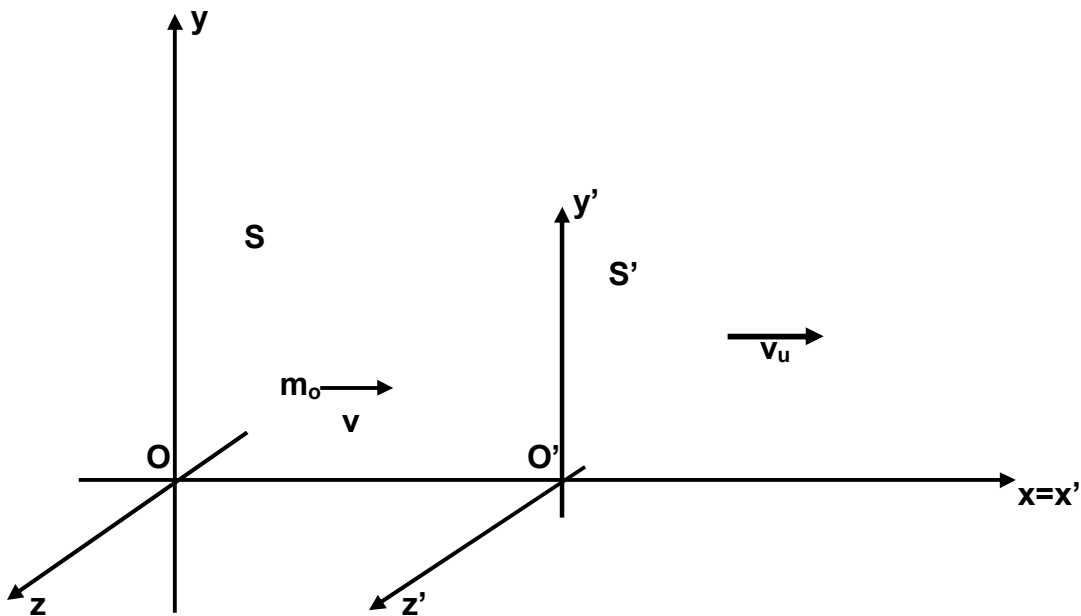

Figure 3. S and S' are inertial reference systems. S is supposed at rest and S' moves with constant speed $v_u$.

Transforming the equation (1.9) we have

$$F = m_o \frac{dv}{dt} + K'(v - v_u) \qquad (1.10)$$

and so

$$k'(v - v_u) = k v$$

Supposing $v_u/v = \beta'$ we have

$$k' = \frac{k}{1 - \beta'} \qquad (1.11)$$

The formula (1.11) represents the theorem of transformation of the resistant coefficient for inertial systems. We observe the resistant coefficient k' depends on the parameter $\beta'$.

### 1.4 The electrodynamic mass

We have considered until now mechanical systems with mass $m_o$ and we have studied the law of motion when they are under the action of a constant mechanical force F. We want now to study the law of motion of the elementary particles which are microphysical systems different from the mechanical systems because they have an elementary electric charge q in addition to mass.
The mechanical systems don't have electric charge and therefore they aren't sensible to electromagnetic forces, the elementary particles on the contrary because of electric charge are sensible to electromagnetic forces and therefore they are electrodynamic systems. Moreover the mass of elementary particles, just because sensible to electromagnetic forces, is substantially different from the mass of mechanical systems sensible to mechanical forces. It is also well known that an accelerated elementary particle emits radiant energy that propagates with the speed of light; mechanical systems don't produce this emission. For all these reasons we call "electrodynamic mass" the mass of elementary particles in order to distinguish it from the "mass" of mechanical systems.
Let us consider therefore a charged elementary particle with electrodynamic mass $m_o$ and null initial speed, accelerated by a constant electric force F.
Under the action of the force F the particle acquires an increasing speed v and at the same time it emits a radiation that propagates with speed c. This radiation originates necessarily from the electrodynamic mass of the particle and therefore we have a transformation of negative electrodynamic elementary mass dm<0 to electromagnetic radiation dW>0, proportional to dm and that propagates with the speed of light c: $dW = -c^2 dm$.
In consequence of this action the electrodynamic mass of the particle decreases while its speed increases.
Supposing the particle is initially at rest and external resistant forces are null (k=0), the equivalent mathematical model of motion is $F = m_o dv/dt$.
The positive elementary work executed by the force F on the particle is
$F ds = m_o v\, dv = dW$ and so by successive calculations we have

$$-c^2 dm = m_o v\, dv$$

$$-c^2 \int_{m_o}^{m} dm = m_o \int_{0}^{v} v\, dv$$

$$-c^2(m - m_o) = \frac{1}{2} m_o v^2$$

$$m = m_o \left[1 - \frac{1}{2} \frac{v^2}{c^2}\right]$$

and putting $\beta = v/c$

$$m = m_o \left(1 - \frac{1}{2}\beta^2\right) \qquad (1.12)$$

The previous formula is the law of real variation of the electrodynamic mass of a charged elementary particle with the speed.
For v=0 we deduce the electrodynamic mass coincides with the electrodynamic mass at rest $m_o$ and for v=c the electrodynamic mass becomes half and so at the speed of light half of the electrodynamic mass of particle is transformed in radiant energy

$$W = -c^2 \int_{m_o}^{\frac{1}{2}m_o} dm = \frac{1}{2} c^2 m_o \qquad (1.13)$$

An accelerated elementary particle emits radiant energy at the expense of the electrodynamic mass and the decrease of the electrodynamic mass is a real physical effect. With this process the work executed by the applied force F on the charged particle gives to this a velocity and at the same time the particle's electrodynamic mass is transformed in radiant energy.
The radiant energy obtained by this process can be considered equivalent to a photon (quantum of electromagnetic energy) with energy $W = m_o c^2/2 = hf$ and with frequency $f = m_o c^2/2h$.
As regards electron we have the following features in frequency and in wavelength (typical magnitudes of $\gamma$ radiations)

$$\lambda = 4{,}9 \times 10^{-6} \; \mu m \qquad \text{and} \qquad f = 6{,}18 \times 10^{19} \; Hz$$

It's right to suppose that at the speed of light also the remaining half the electrodynamic mass is transformed in a second photon with equal energy.
In this way an accelerated elementary particle generates at the speed of light two photons that travel with the speed of light and therefore from a particle with electrodynamic mass at rest $m_o$ we obtain on the whole an energy

$$W = \frac{1}{2} m_o c^2 + \frac{1}{2} m_o c^2 = m_o c^2 \qquad (1.14)$$

We observe by (1.12) moreover that the electrodynamic mass is null when

$$1 - \frac{1}{2}\frac{v^2}{c^2} = 0$$

from which we obtain $v_c = c\sqrt{2} = 1{,}41\ c$.

$v_c$ is the critical velocity of the particle and it is greater than the velocity of light. The electrodynamic mass of a particle can become null by two different mechanisms:

a. production of two equal photons $\gamma$ that travel at the velocity of light
b. production of a "critical particle" when the particle achieves the critical velocity $v_c = 1{,}41\ c$.

The quantity $W = m_o c^2$ represents all the potential energy that it is possible to obtain from the electrodynamic mass at rest $m_o$ of an elementary particle and we call it "intrinsic energy" of the particle.

In this way in the Theory of Reference Frames we have justified completely by the concept of electrodynamic mass the phenomenon of production of couples of photons $\gamma$ from a charged elementary particle and the transformation of electrodynamic mass in energy.

We observe then that for velocities greater than the critical velocity electrodynamic mass is negative and charged elementary particle changes its physical nature becoming a superparticle.

### 1.5 The virtual dynamic mass.

Let's consider a mechanical system with mass $m_o$ at rest in the moving reference frame S' and let's apply in virtual way also to this system the concept of "intrinsic energy" and therefore with respect to the reference frame S' the considered mechanical system has an intrinsic energy $E_o = m_o c^2$.

Since the moving reference frame S' has the speed $v_u$ with respect to the reference frame at rest S, the mechanical system has with respect to S a kinetic energy $E_c = m_o v_u^2 / 2$ and therefore the total energy E of the system with respect to S is

$$E = E_o + E_c = m_o c^2 + \frac{1}{2} m_o v_u^2$$

from which we obtain

$$E = m_o \left(1 + \frac{1}{2}\frac{v_u^2}{c^2}\right) c^2 \qquad (1.15)$$

We can write formally the previous expression $E = m c^2$ where

$$m = m_o \left(1 + \frac{1}{2}\frac{v_u^2}{c^2}\right) \qquad (1.16)$$

and supposing $\beta = v_u / c$ we have

$$m = m_o \left(1 + \frac{1}{2}\beta^2\right) \qquad (1.17)$$

If we call $m_o$ **"static mass"** of the mechanical system at rest the quantity m represents the **"dynamic mass"** for a determined speed $v_u$.

At the speed of light the dynamic mass isn't infinite (like in SR) but equal to $1{,}5m_o$ and this virtual dependence of the mass on speed doesn't put any limit to speed. Besides the variation of mass of a mechanical system with speed isn't real because the virtual increasing of mass is correlated strictly to the kinetic energy of the system and therefore from a point of relativistic view we can affirm that a moving mechanical system with static mass $m_o$ and with speed $v_u$ is equivalent to a system at rest with static mass m. In the Theory of Reference Frames the dynamic mass is therefore a virtual physical concept.

Making a comparison between the virtual relativistic variation of the dynamic mass in TR and the relativistic variation of the transversal mass in SR we can observe, for $v \ll c$, the dynamic mass differs from the transversal mass for terms of fourth and higher order.

## 2. Dynamics of systems with central symmetry

### 2.1 The action at distance in the force fields

The action at distance is the typical force of systems with central symmetry where the point of central symmetry, called pole, is occuped by a central system which generates in surrounding space a force field **F**. Any other secondary system, homogeneous with the central pole and placed in a point of field, is sensible of **F**. The most important remote actions are the force of gravity which causes the fall of bodies, the bigravitational force which causes orbital motions, the force among electric charges which determines the atomic structure.

The experience proves the action at distance between the central system and the secondary system is a force that has an intensity inversely proportional to the square of the distance r between the two systems by a constant $k_a$, called "action constant",

$$F = \frac{k_a}{r^2} \qquad (2.1)$$

The meaning of the action at distance originated from Newton's law of universal gravitation, the meaning of field instead originated in the nineteenth century from the electric and magnetic actions. Between the two meanings there is full compatibility and the action at distance (or remote action) is the force that the central system develops on the secondary system in a force field generated by the central system. The field has a central symmetry and the remote action in theory acts in every point of the field as far as infinity distance where the action is null. The force field is a vectorial field generated by the central system and its remote action acts only when in a point of the field there is a secondary system. The action at distance points out the existence of a potentiality in the force field.

### 2.2 The gravitational motion

If the central system is a pole with static mass $M_o$ the force field generated is the gravitational field which produces a force of attraction at distance, in all the points of space, on a secondary system with static mass $m_o$ ($M_o \gg m_o$).

If the secondary mass $m_o$ is free, not subjected to any tie and not pole, the gravitational force produces the fall of the mass $m_o$ on the surface of the central mass in the direction connecting the two centres of mass. In that case the action constant is $k_a = G M_o m_o$ where G is Newton's constant of universal attraction. The gravitational central force is

$$F = \frac{G M_o m_o}{r^2} \qquad (2.2)$$

and the gravitational field $F_G = F/m_o$ in every point of the field is given by

$$F_G = \frac{G M_o}{r^2} \qquad (2.3)$$

The general law of gravitational motion of the mass $m_o$ is in the described conditions (figure 4)

$$m_o \frac{dv}{dt} + K v = \frac{G M_o m_o}{r^2} \qquad (2.4)$$

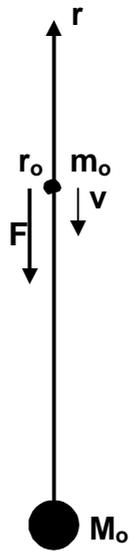

Figure 4. Gravitational motion of the mass $m_o$ in the force field generated by the pole $M_o$.

We derive from the equation (2.4) that the velocity of fall depends on the mass $m_o$ of the secondary system on account of the external resisting forces. Supposing external resistant forces are null (k=0) the velocity of fall is independent of the mass $m_o$ (experience of Newton's tube). In these conditions (k=0) we have

$$\frac{dv}{dt} = \frac{G M_o}{r^2} \qquad (2.5)$$

Suppose that at the initial time t=0 the mass $m_o$ is at the distance $r_o$ with null initial velocity $v(0)=v(r_o)=0$. Moreover it is $v = - dr/dt$ where the minus sign (-) points out that for decreasing values of the distance r in the time (dt>0, dr<0) the modulus of the velocity v is increasing and therefore the acceleration is given by

$$a = \frac{dv}{dt} = -\frac{dv}{dr}\left(-\frac{dr}{dt}\right) = -v \frac{dv}{dr} \qquad (2.6)$$

We have again

$$\frac{v dv}{dr} = -\frac{G M_o}{r^2}$$

and integrating

$$\int_0^v v\, dv = -GM_o \int_{r_o}^r \frac{dr}{r^2}$$

$$\frac{v^2}{2} = -GM_o \left[-\frac{1}{r}\right]_{r_o}^r$$

$$v^2 = \frac{2GM_o}{r_o}\left(\frac{r_o - r}{r}\right)$$

$$v(r) = \sqrt{\frac{2GM_o}{r_o}\left(\frac{r_o - r}{r}\right)} \qquad (2.7)$$

Carrying in diagram the relationship (2.7) we have the graph of velocity of fall (for k=0) (figure 5).

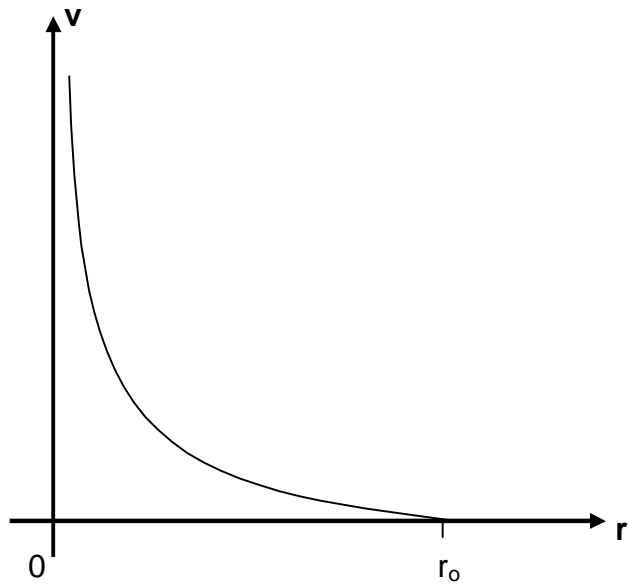

Figure 5. Graph of velocity of fall.

We can observe in figure that in the beginning of the fall and for long part the motion is with good approximation uniformly accelerated in concordance with experimental trial.

In order to move the mass $m_o$ along the negative elementary distance dr the positive elementary work executed by the gravitational force is dW=-Fdr

$$dW = -\frac{GM_o m_o\, dr}{r^2} \qquad (2.8)$$

and therefore the total work executed in order to carry the mass from the distance $r_o$ to the distance $r<r_o$ is

$$W = - G\, M_o\, m_o \int_{r_o}^{r} \frac{dr}{r^2} \qquad (2.9)$$

$$W = G\, M_o\, m_o \frac{r_o - r}{r_o\, r} = k_a \frac{r_o - r}{r_o\, r} \qquad (2.10)$$

It is possible to demonstrate (by 2.7) the work W coincides with the variation of kinetic energy acquired by the mass $m_o$ passing from null velocity to velocity v

$$W = \Delta E_c = E_c(r) - E_c(r_o) = \frac{1}{2} m_o v^2 \qquad (2.11)$$
.

It is well known the potential energy $E_p$ of the gravitational field may be defined

$$E_p(r) = - \frac{G\, M_o\, m_o}{r} \qquad (2.12)$$

and the gravitational force coincides just with the derivative of the potential energy

$$F = \frac{dE_p}{dr} = \frac{G\, M_o\, m_o}{r^2} \qquad (2.13)$$

and therefore

$$W = \Delta E_c = E_p(r_o) - E_p(r) = - \Delta E_p \qquad (2.14)$$

The work executed by the gravitational force on the mass $m_o$ is also equal to the opposite variation of potential energy of the field.

### 2.3 Dynamics of elementary particles in a central field

The central system is constitued by a constant positive electric charge +Q that is the cause of an electric field with central symmetry. An elementary particle with charge -q placed at distance r is subjected to an attractive remote force given by Coulomb's law

$$F = \frac{Q\, q}{4 \pi \varepsilon_o r^2} \qquad (2.15)$$

where the action constant is $k_a = Qq/4\pi\varepsilon_o$.
The law of motion of elementary particle in the field is

$$m_o \frac{dv}{dt} + K v = \frac{k_a}{r^2} \qquad (2.16)$$

where $m_o$ is the particle's electrodynamic mass at rest. Ignoring external resistant forces (k=0 or k≈0) we have

$$m_o \frac{dv}{dt} = \frac{k_a}{r^2} \qquad (2.17)$$

Supposing that at the initial time t=0 the particle is at infinite distance from the central charge ($r_o = \infty$) with null initial velocity $v(t=0) = v(r_o = \infty) = 0$ and is has an intrinsic energy $m_o c^2$, solving the (2.17) we obtain that for any distance r the velocity is

$$v(r) = \sqrt{\frac{Qq}{2\pi\varepsilon_o m_o r}} \qquad (2.18)$$

The velocity of the particle depends on its electrodynamic mass also in the absence of external resistant forces. We know [see First part–paragraph 1.4] that the electrodynamic mass changes with velocity by the law

$$m = m_o \left[ 1 - \frac{1}{2} \frac{v^2}{c^2} \right] \qquad (2.19)$$

from which

$$m = m_o \left[ 1 - \frac{Qq}{4\pi\varepsilon_o m_o r c^2} \right] \qquad (2.20)$$

For $r = \infty$ we have $m = m_o$ and $\Delta m = 0$. At the distance r

$$\Delta m = m - m_o = \frac{-Qq}{4\pi\varepsilon_o c^2 r} \qquad (2.21)$$

Moving the particle from infinite distance to distance r Coulomb's central force executes a positive work W which is equivalent to radiant energy at the expense of the electrodynamic mass

$$W = -c^2 \Delta m = \frac{Qq}{4\pi\varepsilon_o r} \qquad (2.22)$$

As in the electric field the elementary particle -q has a potential energy

$$E_p(r) = -\frac{Qq}{4\pi\varepsilon_o r} \qquad (2.23)$$

the work W is also equal to the opposite variation of potential energy

$$W = -\Delta E_p = -(E_p(r) - E_p(\infty)) \qquad (2.24)$$

where $E_p(\infty) = 0$. In mechanical systems the work executed by the force field is transformed (for k=0) completely in kinetic energy, for elementary particles the work executed by the electric force field is transformed completely in radiant energy at the expense of the electrodynamic mass of particle. At the critical distance

$$r_c = \frac{Qq}{4\pi\varepsilon_o m_o c^2} \qquad (2.25)$$

the electrodynamic mass becomes null (m=0) and it is transformed completely in radiant energy. If the central charge Q and the secondary charge -q have both elementary electric charge equal to the electron we have

$$r_c = \frac{1{,}6^2 \times 10^{-38}}{4 \times 3{,}14 \times 8{,}86 \times 10^{-12} \times 9{,}11 \times 10^{-31} \times 9 \times 10^{16}} = 2{,}8 \times 10^{-15} \text{ m} \quad (2.26)$$

The distance $r_c$ is called "critical distance" because at this distance from the central charge the electrodynamic mass of particle becomes completely null; at greater distances the life of particles is possible, at lower distances elementary particles cannot live in the traditional form and here the life of nucleus begins. At the critical distance $r_c$ the velocity of the electron (with electric charge -e) accelerated in a central field by the action at distance of a positive elementary charge (Q=+e) is

$$v(r_c) = \sqrt{\frac{e^2}{2\pi\varepsilon_o m_o r_c}} = \sqrt{\frac{1{,}6 \times 10^{-38}}{6{,}28 \times 8{,}86 \times 10^{-12} \times 9{,}11 \times 10^{-31} \times 2{,}8 \times 10^{-15}}} = 4{,}25 \times 10^8 \frac{m}{s} \quad (2.27)$$

and therefore we have again the critical velocity is

$$v(r_c) = v_c = c\sqrt{2} = 4{,}25 \times 10^5 \text{ km/s} > c \quad (2.28)$$

The velocity of electron at the critical distance $r_c$ is greater than the velocity of light. We call this particle "critical electron", it is an elementary particle provided with null electrodynamic mass, conventional charge - e, velocity $v_c$>c, null intrinsic energy and nuclear potential energy $E_p(r_c) = - m_o c^2$.
The concept of "electrodynamic mass" of elementary particles confirms Ettore Majorana's intuition who believed that in the nucleus there aren't electrons in traditional form.

### 2.4 The orbital motion

The orbital motion is a motion with central symmetry in which the secondary system, so is it with physical specifications of pole, executes an orbit around the central system. The motion of planets around a star or satellites around a planet are the best known orbital motions.
The central system provided with static mass $M_o$ exerts a gravitational force over the secondary system provided with static mass $m_o$ and this force is given by the force of attraction $F = k_a/r^2 = GM_o m_o/r^2$.
Because of the orbital motion of the secondary system the attraction force can be considered a centripetal force balanced at every time by a centrifugal force $F_c = m_o v^2/r$ where v is the orbital velocity of the secondary system. The balancing of the two forces causes a circular orbital motion with radius at constant curvature

$$r_o = \frac{GM_o}{v^2} \quad (2.29)$$

In reality we know that orbital motions are elliptical and not circular: the elliptical orbit is a consequence of the principle of action and reaction. In fact according to this principle, because the secondary system $m_o$ has physical specifications of pole it generates a force field and this field exerts over the central system an equal and opposite attraction force which moves the central system. As generally $M_o >> m_o$ the action at distance exerted by the central mass $M_o$ over the secondary mass $m_o$ has a greater effect than the action at distance exerted by the secondary mass over the central mass.

Anyway as the central system is free and not still this second action of reaction tends to move the central system (figure 6).

The consequence of this reaction is that the central mass is obliged to move in its turn on a small circular orbit, with small radius with respect to the radius of the circular orbit of the secondary mass, so that the centrifugal force $F'=F_c$ acting on the central mass balances the force of attraction $F''=F$ of the secondary mass on the central mass.

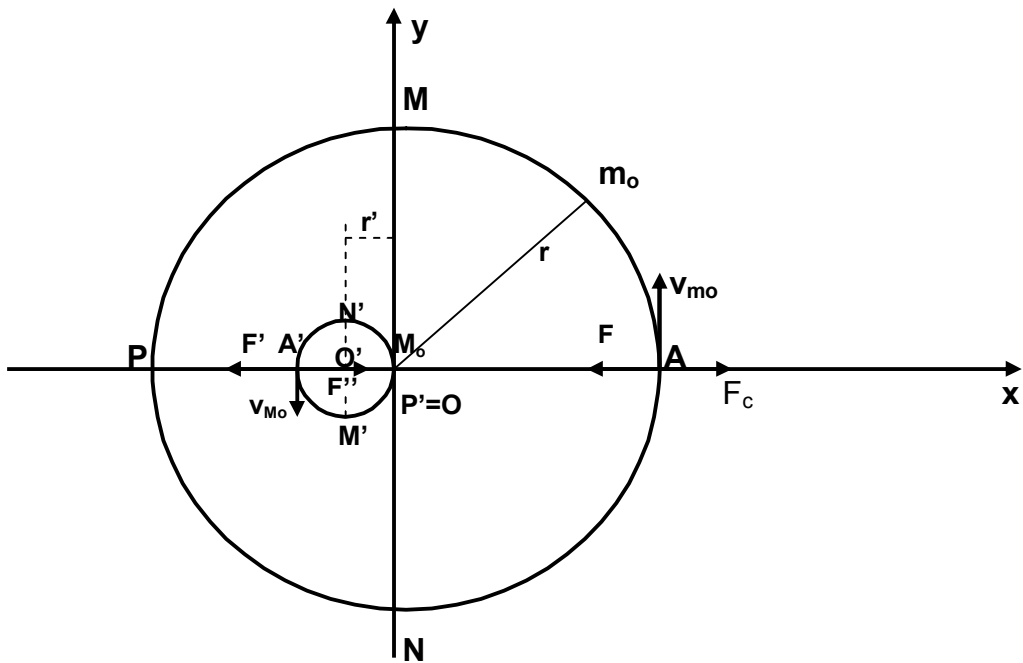

Figure 6. The secondary mass $m_o$ moves around the central mass $M_o$ and for reaction the central mass moves on a small circular orbit.

The secondary mass performs a circular orbit with radius r around the central mass and this performs a circular small orbit with radius r' around the still point O'. The composition of these two motions brings about the elliptical orbit of the mass $m_o$ around the central mass $M_o$ (figure 7).

When $m_o$ is in the point A $M_o$ is in A'. When $m_o$ is in the point P, $M_o$ is in P'=O. If we suppose the central mass is still in the point O', by a simple graphic construction we can observe the secondary mass performs an elliptical motion where the point $A_o$ is the aphelion, the point $P_o$ is the perihelion and the point O' is the focus (fig.7) with $AA_o=A'O'$, $PP_o=P'O'$ and $OA_o^2=OM_o^2+OO'^2$.

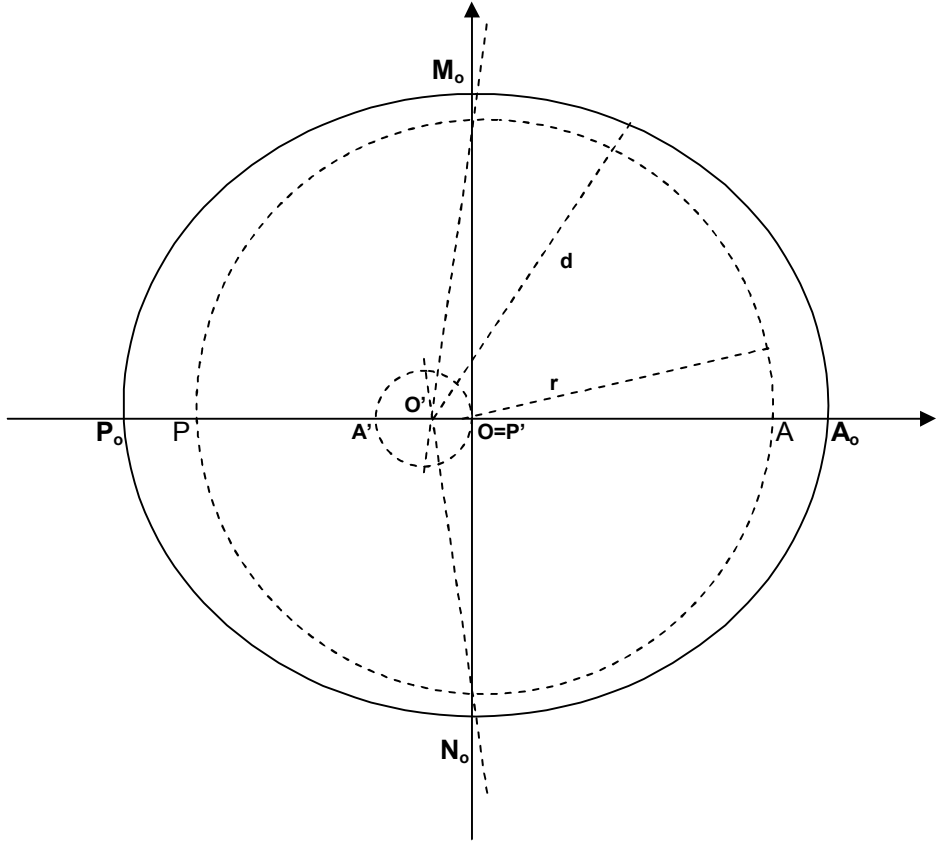

Figure 7. Geometric construction of the orbital motion

In the elliptical motion we observe apparently the central mass is still in the position coinciding with the focus O' of the elliptical orbit.
The eccentricity "e" is the ratio between the radius OO' of the orbit of the central mass (focal distance) and the length $OP_o$ of the greater semiaxis $e = OO'/OP_o$.
If d is the distance between the secondary mass and the central mass in the elliptical orbital motion, making equal the force of attraction and the centrifugal force we have

$$\frac{G M_o m_o}{d^2} = m_o \frac{v^2}{d}$$

$$v = \sqrt{\frac{GM_o}{d}} \qquad (2.30)$$

The velocity is greater in the perihelion $P_o$ and it is smaller in the aphelion $A_o$.
We can deduce the following conclusions:
1. The orbits of the revolving secondary masses are circular with constant radius r
2. The central mass in its turn performs a small circular motion around a still point with constant radius r'
3. If we consider still the central mass in the point O' (focus), in the apparent motion the revolving secondary mass performs an elliptical orbit that obeis Keplero's laws.

The elliptical orbital motion is the result of a field of bigravitational force in which two gravitational forces act at the same time: the gravitational force of attraction of the central mass on the secondary mass (centripetal force) and the gravitational force of attraction of the secondary mass on the central mass (centrifugal force).

In the Theory of Reference Frames[5][6] we have proved the orbital reference systems have the same inertial time of the reference system at rest with origin in O'. This conclusion means the orbital reference systems are inertial reference systems and therefore we can enunciate the following **general principle of inertia**: "a system tends to maintain its state of apparent rest or inertial motion (rectilinear or orbital) until external forces modify the state of the system".
In this way we have fixed a complete inertial equivalence between the rectilinear uniform motion (with k=0) and the orbital motion.

### 3. Relativistic effects in not inertial reference systems

#### 3.1 Relation between time and mass

The reference system S[x,y,z,t,] is supposed at rest and the reference system S'[x',y',z',t'] moves with respect to S with variable velocity **u** and constant acceleration **a₀** (figure 8). The material point P is inside the moving reference frame S' and moves with velocity **v'** with respect to S' and with velocity **v** with respect to S. S' is the privileged reference frame and its own time t' is the inertial time of the examined physical event.

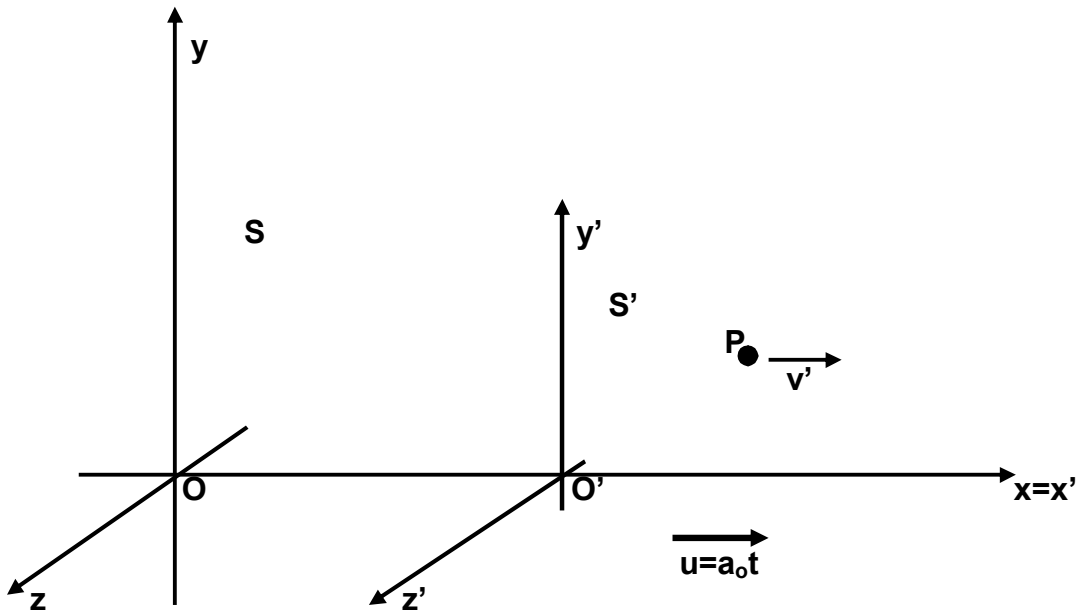

Figure 8. S and S' are not inertial reference systems. S is supposed at rest and S' moves with constant acceleration a₀.

Supposing external resistant forces are null (k'=0, k=0) the law of motion of the point P is F'=m'dv'/dt' with respect to the reference system S' and F=mdv/dt with respect to the reference system S.
The mass m' of the material point is placed under the action of the force F' with respect to the reference system S' and the mass m of the same material point is placed under the action of the force F with respect to the reference system S.
The force F is obtained adding to the force F' the force

$$F_o = m \frac{du}{dt} \qquad (3.1)$$

and as in the considered hypothesis F' and $F_o$ have te same direction and verse we have

$$F = F' + F_o \qquad (3.2)$$

As v=v'+u and du/dt =$a_o$ we have

$$F - F_o = F'$$

$$m \left( \frac{dv}{dt} - \frac{du}{dt} \right) = m' \frac{dv'}{dt'}$$

$$m \frac{d(v - u)}{dt} = m' \frac{dv'}{dt'}$$

$$m \frac{dv'}{dt} = m' \frac{dv'}{dt'}$$

$$m \frac{dv'}{dt'} \frac{dt'}{dt} = m' \frac{dv'}{dt'}$$

$$\frac{dt}{dt'} = \frac{m}{m'} \qquad (3.3)$$

$$dt = \frac{m}{m'} dt' \qquad (3.4)$$

The (3.3) and (3.4) are important relations that link the time and the mass of the material point with respect to the two reference frames. From these two relations it arises that if the two masses are not equal, the two considered reference systems aren't synchronous: in our situation the time t' of the reference frame S', where the physical event happens (privileged reference frame according to the Principle of the Reference), is the inertial time while the time t of the reference frame S is the time of the observer in S. In the Theory of Reference Frames no twin paradox is possible: the time t' is the own real time of the physical event, the time t is the calculated time of laboratory.

### 3.2 The time of electrodynamic systems

For the microphysical systems constitued by charged elementary particles a real variation of electrodynamic mass with the velocity has been proved whether in the systems with linear symmetry or in the systems with central symmetry and therefore moving charged elementary particles have by (3.4) a their own time t' that is different from the time t of the reference frame at rest S. The relativistic variation of time in the Theory of Reference Frames is determined by a variation of electrodynamic mass and no by a law of kinematic transformation. We know

$$m = m' \left( 1 - \frac{1}{2} \frac{u^2}{c^2} \right) \qquad (3.5)$$

where m' is the electrodynamic mass with respect to S' and m is the moving electrodynamic mass with velocity u with respect to S. If the velocity u is constant, as in inertial reference frames, integrating the (3.4) with the coinciding initial time t=t'=0 we have

$$t = t'\left(1 - \frac{1}{2}\frac{u^2}{c^2}\right) \qquad (3.6)$$

with t<t'. In not-inertial reference frames with constant acceleration $u=a_o t$ we have

$$\frac{dt}{dt'} = 1 - \frac{1}{2}\frac{a_o^2 t^2}{c^2} \qquad (3.7)$$

Assuming $k_t = a_o/c\sqrt{2}$ we can write

$$dt' = \frac{dt}{1 - k_t^2 t^2} \qquad (3.8)$$

and integrating with the coinciding initial time t=t'=0

$$t' = \frac{1}{k_t}\ \text{sett tgh}\ k_t t = \frac{1}{2 k_t}\log\frac{1 + k_t t}{1 - k_t t} \qquad (3.9)$$

where $t<1/k_t$. From (3.9) we deduce

$$t = \frac{1}{k_t}\text{tgh}\ k_t t' \qquad (3.10)$$

with t<t'.
The variation of electrodynamic mass causes a variation of the calculated time t for the reference frame at rest S with respect to the inertial time of the moving reference frame S'. This relativistic effect of time is determined for charged elementary particles by a variation of electrodynamic mass and no by a law of kinematic transformation. For the examined event the moving reference frame S' is the privileged reference frame because the material point is and moves in S'.
The time t of the moving charged particle, calculated with respect to the reference frame at rest, passes more slowly (t<t') than its own inertial time t' whether in inertial reference frames (see 3.6) or in not-inertial reference frames (see 3.10). Charged particles with different velocities are different particles with respect to the reference frame at rest. From the physical wiewpoint this means that a charged elementary particle changes its physical nature if its velocity changes. For a particle moving in a force central field at the critical distance the electrodynamic mass and its average life become null with respect to the reference frame at rest. A charged elementary particle in critical conditions becomes a particle with physical features different from a classical elementary particle. Moreover we observe the relativistic effect of time for charged elementary particles produces a contraction of time with respect to the reference frame at rest.

### 3.3 The time of mechanical and electromagnetic systems

**3.3.1** In mechanical systems with linear symmetry and with central symmetry (gravitational systems) there isn't a proved real variation of mass with the velocity whether for inertial reference frames or not-inertial reference frames and therefore S and S' are synchronous because times t and t' are always equal. In fact because the two masses are equal in the two reference systems (m=m') we have dt' = dt and therefore with the coinciding initial time t=t'=0

$$t' = t \quad (3.11)$$

We have proved in the first part (paragraph 1.5) the mechanical systems have a virtual dynamic mass variable with the velocity

$$m = m' \left(1 + \frac{1}{2} \frac{u^2}{c^2}\right)$$

Let us repeat still now that this variable mass is a concept virtual and no real, but using it we have for inertial systems (u=constant)

$$t = t' \left[1 + \frac{1}{2} \frac{u^2}{c^2}\right] \quad (3.12)$$

and for not-inertial systems with constant acceleration $a_o$ ($u=a_o t$)

$$t = \frac{1}{k_t} \text{tg } k_t t' \quad (3.13)$$

with t>t' whether in (3.12) or in (3.13). Let us observe therefore in mechanical systems there is a relativistic effect of time dilation (virtual no real) and this effect for inertial systems is different from the relativistic effect of time in SR for terms of fourth and higher order.

**3.3.2** Electromagnetic and optical systems (electromagnetic waves, light, photons) don't have mass because they are energy systems. Accordingly for these systems there isn't variation of mass and consequently variation of time. All electromagnetic systems have the same inertial time with respect to any reference frame, inertial or not-inertial. We know it is possible to associate photons and electromagnetic waves with a virtual equivalent mass, but in that case all effects of time are virtual and without any real physical meaning.

### 3.4 General equations of space-time transformation for reference frames provided with any velocity

According to that we have explained in this paper and in the ref.[6] we can write the following general equations of space-time transformation that are valid for all inertial and not inertial reference frames (like in fig.8)

$$\begin{cases} x = x' + u_x t \\ y = y' + u_y t \\ z = z' + u_z t \\ dt = m dt'/m' \end{cases} \quad (3.14)$$

and likewise

$$\begin{cases} x' = x - u_x t \\ y' = y - u_y t \\ z' = z - u_z t \\ dt' = m'dt/m \end{cases} \qquad (3.15)$$

The (3.14) and (3.15) allow an important consideration: they prove the time is born when the mass is born. Time without mass cannot be, time without mass would be indeterminate.

For mechanical systems the general equations of space-time transformation become simpler

$$\begin{cases} x = x' + u_x t \\ y = y' + u_y t \\ z = z' + u_z t \\ t = t' \end{cases} \qquad (3.16)$$

and likewise

$$\begin{cases} x' = x - u_x t \\ y' = y - u_y t \\ z' = z - u_z t \\ t' = t \end{cases} \qquad (3.17)$$

that are concordant with the equations of space-time transformation proved in the ref.[6].